\documentclass[twocolumn,aps,amsmath,amssymb]{revtex4-2}
\usepackage[T1]{fontenc}
\usepackage{graphicx}
\usepackage{braket}
\usepackage{hyperref} 
\hypersetup{colorlinks,citecolor=red,urlcolor=blue}

\begin{document}

\title{Guided sampling ans{\"a}tzes for variational quantum computing}

\author{Daniel Gunlycke}
\email{lennart.d.gunlycke.civ@us.navy.mil}
\affiliation{U.S. Naval Research Laboratory, 4555 Overlook Ave SW, Washington, DC 20375, United States}
\author{John P. T. Stenger}
\affiliation{U.S. Naval Research Laboratory, 4555 Overlook Ave SW, Washington, DC 20375, United States}
\author{Andrii Maksymov}
\affiliation{IonQ, 4505 Campus Dr, College Park, MD 20740, United States}
\author{Ananth Kaushik}
\email{kaushik@ionq.co}
\affiliation{IonQ, 4505 Campus Dr, College Park, MD 20740, United States}
\author{Martin Roetteler}
\affiliation{IonQ, 4505 Campus Dr, College Park, MD 20740, United States}
\author{C. Stephen Hellberg}
\affiliation{U.S. Naval Research Laboratory, 4555 Overlook Ave SW, Washington, DC 20375, United States}

\begin{abstract}
Quantum computing is a promising technology because of the ability of quantum computers to process vector spaces with dimensions that increase exponentially with the simulated system size.  Extracting the solution, however, is challenging as the number of quantum gate operations and quantum circuit executions must still scale at most polynomially.  Consequently, choosing a good ansatz—a polynomial subset of the exponentially many possible solutions—will be critical to maintain accuracy for larger systems.  To address this challenge, we introduce a class of guided sampling ans{\"a}tzes (GSAs) that depend on the system interactions and measured state samples as well as a parameter space.   We demonstrate a minimal ansatz for the hydronium cation H$_3$O$^+$ and found that with only 200 circuit executions per structure on the IonQ Aria quantum computer, our calculations produced total energies around the relaxed structure with errors well below $1.59\times10^{-3}$\,Ha, thus exceeding chemical accuracy.
\end{abstract}

\maketitle

\section{Introduction}

The simulation of electronic systems is a promising quantum computing application area.  Electronic systems meet two criteria that combined makes the classical and quantum computing resources required for simulation increase exponentially and polynomially, respectively, with increasing system size.  First, the number of interactions must scale polynomially, or else the system cannot be described on any computer, classical or quantum.  Second, the number of states must scale exponentially, or else there is no need for a quantum computer.  To be able to describe these states using polynomial resources, the state space must be restricted to an ansatz.  The consequence is that the chosen ansatz limits the accuracy of any computed state.  Furthermore, we predict that the expected computational accuracy for a random ansatz will decrease with increasing system size.  To overcome this challenge, we introduce herein a class of GSAs defined by system interactions and measured state samples as well as a parameter space to maintain accuracy as computed systems become larger.  We also present results using a minimal GSA for the hydronium cation H$_3$O$^+$ and found that the accuracy of our calculations, in which sampling was performed on the IonQ Aria quantum computer, well exceeded chemical accuracy.

Energy is a fundamental property in quantum as well as classical mechanics that can be used to predict the time evolution of the states of an isolated system.  In a quantum-mechanical system described by the Hamiltonian $\hat H$, the energy is the expectation value
\begin{equation}
	E\big[\ket{\Psi}\big]=\frac{\braket{\Psi|\hat H|\Psi}}{\braket{\Psi|\Psi}},
	\label{e.1}
\end{equation}
for any state $\ket{\Psi}$ in the representation space $\mathcal F$.  Typically, a main objective in computational physics and chemistry is to find the ground state $\ket{\Psi_\text g}$, i.e. the state with the minimum energy $E_\text g$, and from this state predict a variety of properties from the expectation values of associated observables.  The problem in conventional classical computing is that because the dimension of $\mathcal F$ for many-electron systems increases exponentially with increasing system size, the classical processing unit (CPU) resources needed to compute Eq.\,(\ref{e.1}) also increases exponentially.  With the assistance of a quantum processing unit (QPU) on the other hand, this equation can be computed efficiently to statistical accuracy for all the states that can be efficiently prepared on its quantum register.  Unfortunately, most states cannot be efficiently prepared as the number of quantum gates needed to prepare an arbitrary state would again in general increase exponentially with increasing system size.

\begin{figure}[t]
	\centering
	\includegraphics[width=\columnwidth]{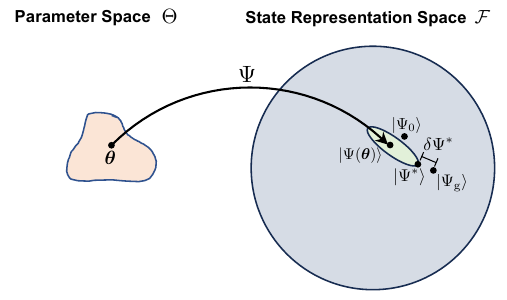}
	\caption{The mapping $\Psi$ from the parameter space $\Theta$ to the state representation space $\mathcal F$ defined by $\boldsymbol{\theta}\mapsto\ket{\Psi(\boldsymbol{\theta})}$ is critical as the ansatz covers only a subset of $\mathcal F$ with a ratio that decreases exponentially with the system size.  To ensure that the distance $\delta\Psi^*$ between the optimized state $\ket{\Psi^*}$ in the ansatz and the sought ground state $\ket{\Psi_\text g}$ is small, the GSA is constructed using measurements of the guiding state $\ket{\Psi_0}$ prepared using system information to be in the neighborhood of $\ket{\Psi_\text g}$.}
	\label{f.1}
\end{figure}
In variational algorithms~\cite{Peruzzo14,McArdle20,Tilly22}, the state preparation challenge is addressed by introducing a mapping $\Psi$ from a parameter space $\Theta$ to the representation space $\mathcal F$.  See Fig.~\ref{f.1}.  We then seek the optimized state $\ket{\Psi^*}$, which is the state in the ansatz (i.e. the image under $\Psi$ of $\Theta$) that has the lowest energy 
\begin{equation}
	E^*=\min_{\boldsymbol{\theta}\in\Theta} E\big[\ket{\Psi(\boldsymbol{\theta})}\big],
	\label{e.2}
\end{equation}
where $\boldsymbol{\theta}$ are parameter vectors in the $d$-dimensional real coordinate space $\mathbb R^d$, in which the parameter space $\Theta$ is a subset.  Because the number of component parameters must not scale exponentially with the system size, $\Psi$ cannot be surjective for large enough systems; rather, the ansatz must be an exponentially small subset of $\mathcal F$.  

There are different classes of ans{\"a}tzes that use various amount of information about the systems of interest.  On one end of the spectrum are pure hardware-efficient ans{\"a}tzes implemented in shallow quantum circuits making maximum use of available QPU resources~\cite{Kandala17,Kandala19,Kokail19,Nakanishi19,Bravo20,Heja23}.  These ans{\"a}tzes are excellent for small systems but are expected to yield inaccurate solutions for larger systems as their locations in $\mathcal F$ are random in the sense that they do not offer any mechanism to hone in on the ground state~\cite{McArdle20,Tilly22}.  To improve solution quality, there are methods to optimize the quantum-gate arrangement~\cite{Kivlichan18,Cincio18,Rattew19,Chivilikhin20,Cincio21,Tkachenko21,Du22,Zhang22} or protect one or more symmetries of the calculated systems~\cite{Barkoutsos18,Ganzhorn19,Otten19,Gard20,Seki20}.  On the other end of the spectrum are the quantum adiabatic evolution ans{\"a}tzes~\cite{Farhi2001} that converge to the ground state but also require the deepest circuits.  A compromise between hardware-efficient ans{\"a}tzes and converging ans{\"a}tzes are physically-motivated ans{\"a}tzes, including unitary-couple-cluster (UCC) ans{\"a}tzes~\cite{Peruzzo14,Yung14,McClean16,OMalley16,Shen17,Barkoutsos18,Kivlichan18,Hempel18,Harsha18,Romero18,Dallaire19,Lee19,Matsuzawa20,Sokolov20,Nam20,Grimsley20,Bauman21,Motta21,Anand22}.  These ans{\"a}tzes consider up to a given number of excitations from the Hartree--Fock (HF) state but otherwise do not use system-specific information.  To reduce the circuits depth for UCC ans{\"a}tzes, one can apply adaptive ans{\"a}tzes that optimize excitation operators~\cite{Grimsley19,Tang21,Yordanov21} at the expense of additional circuit executions.  There are also other types of ans{\"a}tzes for quantum subspace methods~\cite{Endo21,Epperly22,Motta24} that use subspace expansions~\cite{McClean17,Colless18,Urbanek20,Motta21,Endo21,Epperly22,Motta24,Umeano25}, including ones that use multi-state contraction subspace search~\cite{Parrish19,Nakanishi19,Huggins20,Heja23}, Krylov subspaces~\cite{Parrish19arXiv,Stair20,Baker21,Bharti21,Cohn21,Jamet21,Jamet22,Baker24,Yoshioka25}, and sample-based subspaces~\cite{Robledo-Mereno25}.  These diagonalization methods circumvent any convergence issues, including barren plateaus~\cite{McClean18,Grant19}, that can arise in iterative optimization methods.  They also address mismatches present in many ans{\"a}tzes for the variational quantum eigensolver algorithm between the number of parameters that can be processed on the QPU and the CPU.

\section{Results}

\subsection{Guided sampling ansatz}

\begin{figure}[t]
	\centering
	\includegraphics[width=\columnwidth]{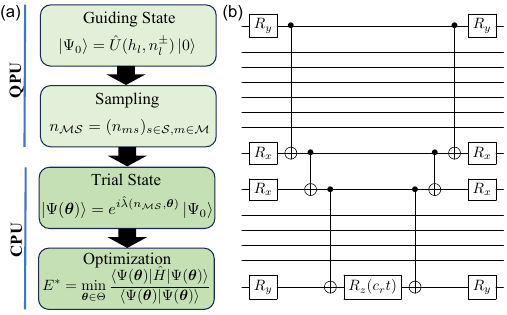}
	\caption{The implementation of a GSA.  (a) The calculation workflow as given by the CVQE algorithm executed sequentially on the QPU and CPU.  The ansatz is defined by the system-dependent operator $\hat U$ introduced in the guiding state and the sample- and parameter-dependent operator $\hat\lambda$ in the trial state.  (b) Circuit component for the Pauli tensor product term $c_rt\,\hat\sigma_x\otimes\hat\sigma_0^{\otimes 6}\otimes\hat\sigma_y^{\otimes 2}\otimes\hat\sigma_0^{\otimes 4}\otimes\hat\sigma_x$ with operators from left to right acting on the qubits represented from the top to bottom line.  The operators $R_x$ and $R_y$ rotate the state around the x- and y-axis, respectively, by $\pi/2$ and the operator $R_z$ around the z-axis by $c_rt$.}
	\label{f.2}
\end{figure}
An alternative approach to address the parameter mismatch is provided by the cascaded variational quantum eigensolver (CVQE) algorithm~\cite{Gunlycke2024,Stenger2023,Stenger2024}.  In this algorithm, one creates a guiding state $\ket{\Psi_0}=\hat U\ket{0}$ on the QPU defined by a $\boldsymbol{\theta}$-independent operator $\hat U$ and the register product state $\ket{0}$ that has all qubits in the zero state.  

The $\boldsymbol{\theta}$-dependence is instead introduced through a separate linear operator $\hat\lambda(\boldsymbol{\theta})$ that defines the trial state $\ket{\Psi(\boldsymbol{\theta})}=e^{i\hat\lambda(\boldsymbol{\theta})}\ket{\Psi_0}$.  The advantage of this construction is that the quantum state sampling and optimization have been separated such that the former sampling of $\ket{\Psi_0}$ can be performed on the QPU before the optimization of $\ket{\Psi(\boldsymbol{\theta})}$ commences on the CPU.  See Fig.~\ref{f.2}(a).

Because of the cascaded form of the sampling and optimization processes, we can make the trial state and concomitantly the ansatz dependent on the measured samples $n_{\mathcal M\mathcal S}=(n_{ms})_{s\in\mathcal S,m\in\mathcal M}$ through the mapping $\hat\lambda=\hat\lambda(n_{\mathcal M\mathcal S})$ from $\Theta$ to the set of linear operators on $\mathcal F$, where $n_{ms}$ is the measured outcome of shot $s$ in the set $\mathcal S$ with the probability mass function $\operatorname{P}[\hat R_m\Psi_0\!\mapsto\! n_{ms}] = \big|\!\braket{\Psi_0|\hat R_m^\dagger|n_{ms}}\!\big|^2$, for all measurement operators $\hat R_m$ indexed over the measurement set $\mathcal M$.  The total number of measurements $N_\text m=|\mathcal M||\mathcal S|$ is generally several orders of magnitude lower than in the original variational quantum eigensolver algorithm.

Moreover, as $\hat U$ is independent of $\boldsymbol{\theta}$, we can focus the available quantum resources on creating and measuring a system-dependent guiding state designed to be close to the ground state.  Specifically, we let $\hat U=\hat U(h_{\mathcal L},n_{\mathcal L}^\pm)$, for all interactions indexed over the set $\mathcal L$, where $h_{\mathcal L}=(h_l)_{l\in\mathcal L}$ and $n_{\mathcal L}^\pm=(n_l^\pm)_{l\in\mathcal L}$ are families of coefficients and tuples, respectively, defining the system Hamiltonian
\begin{equation}
	\hat H(h_{\mathcal L},n_{\mathcal L}^\pm)=\sum_{l\in\mathcal L} h_lC_{n_l^+}^\dagger C_{n_l^-},
	\label{e.3}
\end{equation}
where $C_n^\dagger$ are configuration creation operators that generate the Fock states $\ket{n}=C_n^\dagger\ket{0}$.  A GSA is then any map generated by $\Psi=\Psi(\hat U,\hat\lambda)$ defined by the trial state
\begin{equation}
	\ket{\Psi(\boldsymbol{\theta})}=e^{i\hat\lambda(n_{\mathcal M\mathcal S},\boldsymbol{\theta})}\hat U(h_{\mathcal L},n_{\mathcal L}^\pm)\ket{0}.
	\label{e.4}
\end{equation}

The goal is to choose the operators $\hat U$ and $\hat\lambda$ such that the guiding state $\ket{\Psi_0}$ and the optimized state $\ket{\Psi^*}$ are as close as possible to the ground state $\ket{\Psi_\text g}$ given existing QPU and CPU resource constraints.  A measure of closeness for any state $\ket{\Psi}\in\mathcal F$ is the 2-norm
\begin{equation}
	\delta\Psi=\sqrt{2\big(1-\operatorname{Re}\braket{\Psi_\text g|\Psi}\big)},
	\label{e.5}
\end{equation}
for the state deviation $\ket{\delta\Psi}=\ket{\Psi}-\ket{\Psi_\text g}$.  See Fig.~\ref{f.1} for $\ket{\Psi^*}$ specifically.  From this norm also follows the energy deviation
\begin{equation}
	\delta E=\big(E_\delta-E_\text g\big)\big(\delta\Psi\big)^2,
	\label{e.6}
\end{equation}
where $E_\delta$ is the energy of $\ket{\delta\Psi}$.

\subsection{Guiding state}

The ideal guiding state is one that approaches the ground state by adiabatic state evolution~\cite{vanDam2001,Smelyanskiy2002,Farhi2001} from the solution $\ket{\Phi_0}$ of a model system $\hat H_0$.  This can be achieved by choosing
\begin{equation}
	\hat U(h_{\mathcal L},n_{\mathcal L}^\pm)=\mathcal T\exp\bigg\{-\frac{i}{\hbar}\int_0^T\hat H\Big(\frac{t}{T}\Big)dt\bigg\}\hat U_0,
	\label{e.7}
\end{equation}
where $\mathcal T$ is the operator that orders the Hamiltonians such that $\mathcal T\hat H(\eta)\hat H(\eta')=\hat H(\eta)\hat H(\eta')$ for all $\eta\ge\eta'$ and $\mathcal T\hat H(\eta)\hat H(\eta')=\hat H(\eta')\hat H(\eta)$ for all $\eta<\eta'$, $\hat H(\eta)=\hat H_0(1-\eta)+\eta\hat H$ is a parameterized Hamiltonian for all $\eta\in[0,1]$, and $\hat U_0$ is a unitary operator that maps $\ket{0}$ to $\ket{\Phi_0}$.  For many physical and chemical systems, the Hartree--Fock method provides a reasonable model system with a solution $\ket{\Phi_0}$ easily prepared on the quantum register by applying $\text{X}$ gates to those qubits that represent spin-orbitals occupied by electrons.  From the adiabatic theorem, it follows for any given state deviation tolerance $\delta$ that $\delta\Psi_0\le\delta$ if $T\ge40\delta^{-1}\lambda^{-2}\max\big(||\hat V||,\lambda^{-1}||\hat V||^2\big)$, where $\lambda$ is the minimum spectral gap between the ground and first excited state, $\hat V=\hat H-\hat H_0$, and $||\cdot||$ is the 2-norm~\cite{Duan20}.

Without a practical way to implement Eq.\,(\ref{e.7}) directly on the QPU, we first need to approximate it.  One approximation is
\begin{equation}
	\hat U(h_{\mathcal L},n_{\mathcal L}^\pm)\approx\exp\!\!\bigg[\frac{-i\hat H\Delta T}{2\hbar}\bigg]\!\prod_{k=1}^{K-1}\!\exp\!\!\bigg[\frac{-i\hat H\big(1-\frac{k}{K}\big)\Delta T}{\hbar}\bigg]\hat U_0,
	\label{e.8}
\end{equation}
which is obtained by applying the trapezoidal rule with $K$ equally sized steps $\Delta T=T/K$, followed by the generalized Trotter formula~\cite{Trotter59,Suzuki76}, where the factor for $\hat H_0$ has been dropped as it only generates an irrelevant global phase shift when applied to $\ket{\Phi_0}$.  The number of steps $K$ should ideally be large such that $\Delta T\ll\lambda^{-1}$ to let the state $\ket{\Phi_0}$ adiabatically evolve to $\ket{\Psi_0}$ without introducing excitations.  For large $K$, we have $\delta\Psi_0\propto K^{-1}$ from a closely related Trotter approximation~\cite{Suzuki76}.  Furthermore, as the number of gate operations $N_\text g$ needed to implement the operator in Eq.\,(\ref{e.8}) is proportional to $K$, we expect that the state deviation norm $\delta\Psi_0\propto N_\text g^{-1}$.

\subsection{Trial state}

Once the guiding state $\ket{\Psi_0}$ has been prepared and measured, we can use the produced samples $n_{\mathcal M\mathcal S}$ in our choice of $\hat\lambda(n_{\mathcal M\mathcal S},\boldsymbol{\theta})$.  The minimal measurement set contains only the standard measurement basis with the measurement operator given by the identity operator.  One special choice using this minimal set is the operator
\begin{equation}
	\hat\lambda(n_{\mathcal S},\boldsymbol{\theta})=\sum_n \lambda_n(n_{\mathcal S},\boldsymbol{\theta})\ket{n}\!\bra{n}
	\label{e.8-9}
\end{equation}
that is defined by the parametric equation
\begin{equation}
	\lambda_n(n_{\mathcal S},\boldsymbol{\theta})=\Bigg\{
	\begin{array}{cr}
		-i\ln\frac{\theta_n}{\braket{n|\Psi_0}}, & \quad\text{for }n\in\mathcal N_\mathcal S,\\
		i\infty, & \quad\text{for }n\notin\mathcal N_\mathcal S,
	\end{array}
	\label{e.9}
\end{equation}
where $\mathcal M$ has been dropped for notational brevity, $\theta_n$ are complex parameters and $\mathcal N_\mathcal S$ is a sample-dependent set of Fock states.  Although one could choose the latter set to be the straightforward set $\{n_s\}_{s\in\mathcal S}$, we have found that it works better in practice to use
\begin{equation}
	\mathcal N_\mathcal S=\Big\{n\,:\,n\in\mathcal F, n'\in\{n_s\}_{s\in\mathcal S}, \braket{n|\hat H|n'}\ne0\Big\},
	\label{e.10}
\end{equation}
as this choice necessarily accounts for all interactions acting on each measured shot.  In any case, the trial state defined by Eq.\,(\ref{e.9}) is
\begin{equation}
	\ket{\Psi(\boldsymbol{\theta})}=\sum_{n\in\mathcal N_\mathcal S}\theta_n\ket{n}
	\label{e.10-1}
\end{equation}
and associated with this trial state is the minimal GSA
\begin{equation}
	\Psi(\Theta)=\operatorname{span}\!\big\{\!\ket{n}\!\big\}_{n\in\mathcal N_\mathcal S},
	\label{e.10-2}
\end{equation}
which is also a Fock subspace.

Because the number of measurements $N_\text m=|\mathcal S|$ must not increase exponentially with the system size and the Hamiltonian is limited to one- and two-body interactions, the dimension of the GSA increases at most polynomially.  Thus, we can calculate all the Hamiltonian matrix elements on $\Psi(\Theta)$ and solve the eigenvalue equation $\hat H\ket{\Psi_\lambda}=E_\lambda\ket{\Psi_\lambda}$ by direct diagonalization.  The optimized energy $E^*$ in Eq.\,(\ref{e.2}) is then the lowest eigenvalue $E_{\lambda^*}$ and the optimized state $\ket{\Psi^*}$ the associated eigenstate $\ket{\Psi_{\lambda^*}}=\sum_{n\in\mathcal N_\mathcal S}\theta_n^*\ket{n}$, where $\theta_n^*$ are the optimized parameters.

\subsection{Demonstration}

\begin{figure}[t]
	\centering
	\includegraphics[width=\columnwidth]{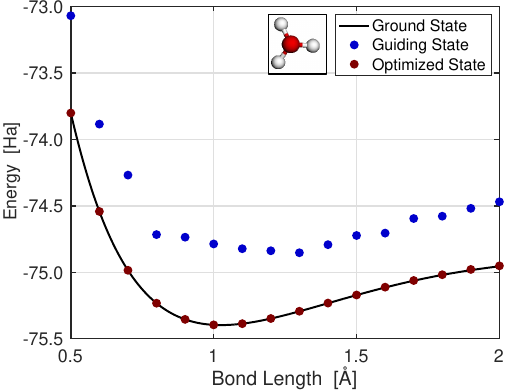}
	\caption{The energy as a function of bond length for the hydronium cation H$_3$O$^+$ with point group symmetry $C_{3v}$ and $14$ valence spin orbitals.  The guiding states $\ket{\Psi_0}$ and their energies were calculated classically for $K=1$, $\varepsilon=0.11$\,Ha, and $T=2\hbar$.  The same guiding states were also generated and sampled on the IonQ Aria quantum computer with $|\mathcal S|=200$ shots for each bond length.  The GSAs were then formed using the measured samples and the trial states $\ket{\Psi(\boldsymbol{\theta})}$ and their energies optimized classically.  The ground state energies are the lowest eigenvalues obtained classicially by exact diagonalization of the full-configuration-interaction Hamiltonian.}
	\label{f.3}
\end{figure}
To test the minimal GSA obtained from Eqs.\,(\ref{e.8})--(\ref{e.10-2}), we performed calculations of the hydronium cation H$_3$O$^+$ shown in the Fig.~\ref{f.3} inset using the IonQ Aria quantum computer.  This cation forms when hydrogen ions (i.e. protons) bond to water molecules and is thus present in acidic water.  Understanding this cation is important for electrochemistry and could help in designing additives to mitigate corrosion in acidic environments.

To limit the impact of quantum decoherence, we set an allowance for the number of gate operations $N_\text g$ in the quantum circuits and chose to use only a single step, $K=1$, in the adiabatic evolution.  While this is an extreme approximation of adiabatic evolution, operating in the small $K$ limit can still work in practice as the primary purpose of the QPU calculations is to identify the most important Fock states through the obtained measurement samples.  The sampling is affected by errors in the magnitudes of the projections of the guiding state on the Fock states in the chosen measurement basis (The sensitivity of the guiding state to such errors needs investigation but is outside the scope of this work).  However, the sampling is unaffected by the corresponding phase errors, as the phases of the optimized trial state are determined during the subsequent CPU calculations.

Even as we chose a single step in the adiabatic evolution, we still needed another way to control $N_\text g$ and chose to neglect interactions with magnitudes smaller than a given $\varepsilon$.  The unitary operator then became $\hat U(h_{\mathcal L},n_{\mathcal L}^\pm)\approx \hat U_1^\varepsilon(h_{\mathcal L},n_{\mathcal L}^\pm)=e^{-i\hat H^\varepsilon T/2\hbar}\hat U_0$, where the Hamiltonian $\hat H^\varepsilon$ is in the form of Eq.\,(\ref{e.3}) but with the coefficients $h_l^\varepsilon$ given by $h_l^\varepsilon=h_l$ for $|h_l|\ge\varepsilon$ and zero otherwise.  To implement this operator, we first expanded the Hamiltonian as $\hat H^\epsilon=\sum_{r\in\mathcal R}c_r^\epsilon\bigotimes_{q\in\mathcal Q}\hat\sigma_{r_q}$, where $\mathcal R=(0,x,y,z)^Q$ and $\mathcal Q=\{1,...,Q\}$ for a QPU register with $Q$ qubits, $c_r^\epsilon$ are Pauli coefficients, and $\hat\sigma_{r_q}$ are identity or Pauli operators, depending on $r_q$.  Then, we applied the first-order generalized Trotter approximation~\cite{Suzuki76}, which yielded $\hat U_1^\varepsilon(h_{\mathcal L},n_{\mathcal L}^\pm)=\hat U(c_{\mathcal R})\hat U_0\approx\big[\prod_{r\in\mathcal R}\hat U_r(c_r)\big]\hat U_0$, where the family $c_{\mathcal R}=(c_r)_{r\in\mathcal R}$ and the operator $\hat U_r(c_r)=e^{-i(c_rT/2\hbar)\bigotimes_{q\in\mathcal Q}\hat\sigma_{r_q}}$.  Next, we generated for each operator $\hat U_r(c_r)$ a circuit component of the form shown in Fig.~\ref{f.2}(b), put the circuit components together, and performed circuit optimization.  The samples obtained from the guiding state were then used to perform the subsequent parameter optimizations on a classical computer.  The produced many-electron energies of the guiding state and the optimized trial state are shown in Fig.~3.

\begin{figure}[t]
	\centering
	\includegraphics[width=\columnwidth]{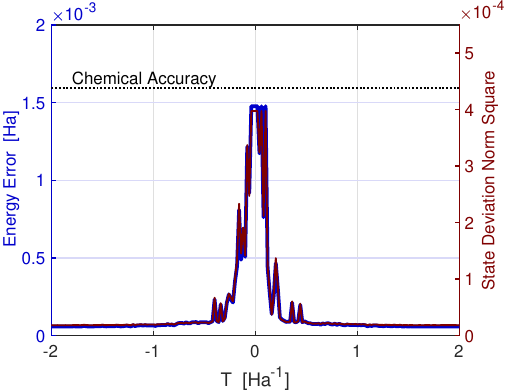}
	\caption{The energy error $\delta E^*$ and the state deviation norm square $(\delta\Psi^*)^2$ for the optimized trial state for H$_3$O$^+$ with the bond length $1.0$\,\AA~as a function of $T$ in atomic units ($\hbar=1$).  The close match between the curves confirms that we can use either $\delta E^*$ or $(\delta\Psi^*)^2$ as a measure of accuracy.  The peak around $T=0$ is a result of the guiding states $\ket{\Psi_0}$ being close to the Hartree--Fock state $\ket{\Phi_0}$, causing the sampling to produce few measurement outcomes corresponding to other Fock states, thereby limiting the size of the parameter space on which the energy minimization occurs.  For the same reason, the observed spikes, which originates from statistical noise from the limited number of shots, in this case $|\mathcal S|=200$, are larger around this peak.}
	\label{f.4}
\end{figure}
For small $K$, it is not immediately clear how the parameter $T$ should be chosen.  If $|T|$ is small, then the first terms in the Taylor expansions of the factors in Eq.\,(\ref{e.8}) would dominate, which means that the guiding state would have large projections onto the Hartree--Fock state $\ket{n_\text{HF}}$ and those states immediately coupled to it by the Hamiltonian.  This further means that a large number of measurements would be needed to obtain a reasonable size of the sample set and concomitantly the GSA.  On the other hand, if $|T|$ is large, the step size $\Delta T$ in Eq.\,(\ref{e.8}) would also be large, which would cause non-adiabatic excitations.  We found that the onset of excitations in our calculations for the bond length $1.0$\,\AA~was just above $|T|/\hbar=2.0$\,Ha$^{-1}$, which is also approximately the inverse Hartree--Fock energy gap between the ground state and the first excited state.

Figure~\ref{f.4} shows the dependence of the energy error and the state deviation of the optimized trial state as a function of $T$ below the excitation threshold.  These curves confirm the expectation that the error increases significantly for $|T|$ small because of the limited number of shots (Not shown is another set of identical calculations with the number of shots increased from 200 to 10,000 that yielded peaks that were much narrower and barely visible).  Moreover, we find that for sufficiently large $|T|$ below the excitation threshold, we need not be concerned with the observed spikes that originate from the statistic noise from the limited number of samples.  Furthermore, from the close match between the two curves, we find that $E_\delta$ in Eq.\,(\ref{e.6}) for the optimized state is almost independent of $T$, which indicates that the optimization is largely able to overcome the sampling error caused by the rough guiding state approximation.  This finding is also consistent with the energy data sets shown in Fig.~\ref{f.3} for the guiding state and optimized state.

\begin{figure}[t]
	\centering
	\includegraphics[width=\columnwidth]{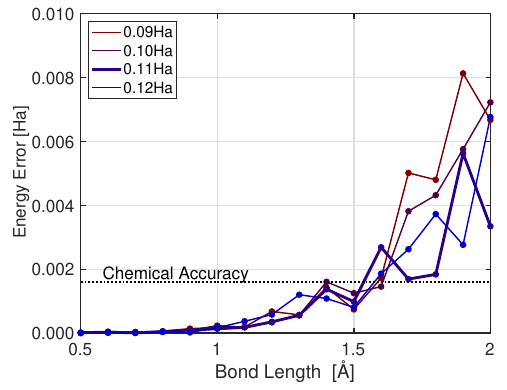}
	\caption{The energy error $\delta E^*$ as a function of bond length for H$_3$O$^+$ for different energy cutoffs $\varepsilon$.  The guiding states $\ket{\Psi_0}$ were generated and sampled on the IonQ Aria quantum computing with $T=2\hbar$ and $|\mathcal S|=200$ shots.  The GSAs were then formed using the measured samples and the trial states $\ket{\Psi(\boldsymbol{\theta})}$ and their energies optimized classically.  The resulting energy errors for the optimized states are at or below chemical accuracy $1.59\times10^{-3}$\,Ha for bond lengths at or below $1.5$\,\AA.  The error increases with increasing bond length as the number of Fock states needed to approximate the ground state to within a given tolerance also increases.  As the impact of quantum decoherence on $\ket{\Psi_0}$ increases with decreasing energy cutoff, as the circuit depth increases, there is an optimum energy cutoff that depends on the quantum hardware.  In our calculations, the optimum energy cutoff $\varepsilon=0.11$\,Ha is highlighted by the thicker curve.}
	\label{f.5}
\end{figure}
Another guiding state parameter that can be tuned is the cutoff energy $\varepsilon$.  If $\varepsilon$ is small, the circuit depth is large and quantum decoherence will generally increase the state deviation norm $\delta\Psi_0$.  If $\varepsilon$ is large, a large number of interactions will be cut from the generation of the guiding state, defeating the purpose of the GSA.  Consequently, there is a optimal middle ground that minimizes the error, as can be seen in Fig.~\ref{f.5}, presenting the energy error of the optimized state for different $\varepsilon$.

Figure~\ref{f.5} also shows that our calculations of the total many-electron energy exceed chemical accuracy (i.e., the energy error is below $1.59\times10^{-3}$\,Ha) for all bond lengths at or below $1.5$\,\AA.  Using the GSA presented herein, we only needed $N_\text m=|\mathcal M||\mathcal S|=200$ quantum circuit executions per atomic structure.  This is over $8$ orders of magnitude less than needed by the original variational quantum eigensolver algorithm with $N_\text m=|\mathcal I||\mathcal M||\mathcal S|$ circuit executions per atomic structure, where the number of iterations is typically at least $|\mathcal I|\sim10^2$, the number of measurements generally $|\mathcal M|\sim(2Q)^4$, and the number of shots typically $|S|\sim10^3$, thus requiring tens of billions of circuit executions for the hydronium cation H$_3$O$^+$.

\section{Discussion}

\begin{figure}[t]
	\centering
	\includegraphics[width=\columnwidth]{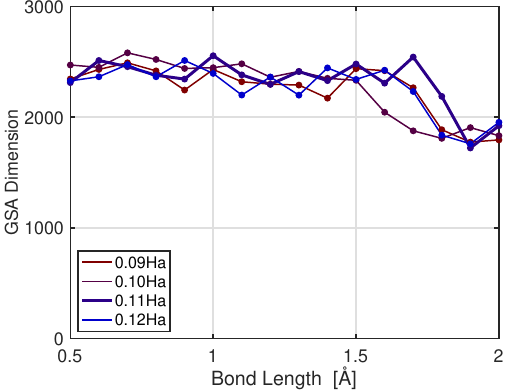}
	\caption{The dimension of the minimal GSA $\Psi[\Theta]$ as a function of bond length for H$_3$O$^+$ for different energy cutoffs $\varepsilon$.  The guiding states $\ket{\Psi_0}$ were generated and sampled on the IonQ Aria quantum computing with $T=2\hbar$ and $|S|=200$ shots.  The GSAs were then formed using the measured samples.  Except for a slight drop in the dimension for large bond lengths, there is otherwise to clear dependence of the dimension on either the bond length or the energy cutoff.}
	\label{f.6}
\end{figure}
Importantly, our calculations exceed chemical accuracy even as the dimension of the minimal GSA $d$ shown in Fig.~\ref{f.6} is almost an order of magnitude smaller than the dimension of the Fock space $D=2^Q=16,384$.  Although our results are promising, it remains unclear how the accuracy depends on the system and sample sizes.  While leaving this for future work, we do expect that the optimized state would generally be more accurate than the guiding state, for which an upper bound of the error could in principle be obtained.

The approach described herein also lets us tune both the needed QPU and CPU resources.  As described above, the QPU circuit size $N_\text g$ can be controlled by the guiding state through the applied approximations, including the number of steps $K$ and the cutoff energy $\varepsilon$.  The number of QPU measurements $N_\text m$ can be controlled by the cardinality of the measurement basis set $|\mathcal M|$ and the sample set $|\mathcal S|$.  For our direct optimization method, the needed CPU resources depends on the dimension $d=\dim \Psi[\Theta]$, which can also be controlled by $|\mathcal S|$.  Moreover, if the $d$ is too large, one can also consider using the sample multiplicity to trim the size.  All information about multiplicity contained in $n_{\mathcal S}$ was dropped by the definitions in Eq.\,(\ref{e.9}) and Eq.\,(\ref{e.10}).  One could use a conventional iterative optimization process if convergence into a global minimum is expected, which should also reduce the CPU resources.

\section{Methods}

\subsection{H$_3$O$^+$ encoding}

For the H$_3$O$^+$ calculations, we considered $16$ atomic structures with O-H bond lengths uniformly spaced between $0.5$\,\AA~and $2.0$\,\AA.  We required that all the structures satisfied the symmetry transformations of point group $C_{3v}$ and had the fixed distance 0.372\,\AA~between the O nucleus and the plane formed by the three H nuclei.  For each structure, we considered the 16 spin-orbitals formed by the spin set $\{\alpha,\beta\}$ and the atomic orbital set $\{1s_1^\text {H},1s_2^\text {H},1s_3^\text {H},1s^\text {O},2s^\text O,2p_x^\text O,2p_y^\text O,2p_z^\text O\}$ and calculated the Hamiltonian coefficients from the one- and two-electron integrals obtained from the Born--Oppenheimer approximation using the minimal basis STO-3G.  We then performed spin-restricted Hartree--Fock calculations within the frozen core approximation and generated for each structure the families $h_{\mathcal L}$ and $n_{\mathcal L}^\pm$ for the produced one-electron Hartree--Fock basis.  This basis comprises the 14 valence spin-orbitals formed by the elements of $\{\alpha,\beta\}$ and the converged molecular orbitals set, whose elements are linear combinations of the 7 valence atomic orbitals in $\{1s_1^\text {H},1s_2^\text {H},1s_3^\text {H},2s^\text O,2p_x^\text O,2p_y^\text O,2p_z^\text O\}$.  The Hartree--Fock calculations also produced the initial Hartree--Fock Hamiltonian $\hat H_0=\hat H_\text{HF}$ and Fock state $\ket{\Phi_0}=\ket{n_\text{HF}}$.  Applying the Jordan--Wigner mapping~\cite{Jordan28}, the latter is the basis state of the QPU register composed of $Q=14$ qubits with a `1' for each of the 8 valence spin-orbitals occupied by electrons and `0' for each of the remaining 6 unoccupied valence spin-orbitals.

\subsection{QPU execution}

For this study, we utilized the IonQ Aria QPU~\cite{Chen24}, a system of 25 trapped $^{171}$Yb$^+$ ions, in which the qubit states are encoded within two hyperfine states that originate from the degenerate ground state of these ions.  The ions are sourced via laser ablation, selectively ionized, and loaded into a surface linear Paul trap in a compact vacuum package.  Control is achieved using 355 nm laser pulses that drive two-photon Raman transitions, enabling arbitrary single-qubit rotations and M\o{}lmer--S\o{}rensen (MS) two-qubit entangling gates~\cite{Sorensen00}.  At the time of execution, these entangling gates had a median direct randomized benchmarking (DRB)~\cite{Proctor19} fidelity of 99.88\% with gate times around 650\,$\mu$s.

Prior to execution, our circuits were optimized in two stages.  First, we minimized the two-qubit gate count by optimizing the Pauli term ordering and their CNOT decompositions~\cite{Goings23}.  Second, these circuits were passed to IonQ’s compiler to be further optimized and transpiled into the native gate set of MS gates and single-qubit rotations.  The produced optimized circuits were executed on the IonQ Aria to collect the required samples.

\section{Data availability}

The data presented in this paper are available from the corresponding authors upon reasonable request.

\section{Code availability}

The software codes used for this project have not been approved for public release.

\bibliographystyle{apsrev4-2}

\section{Acknowledgments}

This work has been supported by the Office of Naval Research (ONR) through the U.S. Naval Research Laboratory (NRL) Base Program.  Andrii Maksymov, Ananth Kaushik, and Martin Roetteler are employees and equity holders of IonQ, Inc.

\section{Author contributions}

D.G., J.P.T.S., and C.S.H. conceived the project.  D.G. secured support, oversaw the project, and wrote the paper with input from all authors.  J.P.T.S. generated the system data, wrote the code, and performed the classical optimizations.  A.M. performed the quantum circuit optimizations.  A.K. oversaw the QPU executions and assembled the samples.  M.R. reviewed the QPU calculations.

\section{Competing interests}

The authors declare no competing interests.

\end{document}